\documentclass[aps,prl,twocolumn]{revtex4}
\usepackage{graphicx}
\usepackage{amsmath}
\usepackage{amssymb}
\usepackage{mathptmx} 

\begin{document}
\title{Weyl points and line nodes in gapless gyroid photonic crystals}
\author{Ling Lu, Liang Fu, John D. Joannopoulos and Marin Solja\v{c}i\'{c}}
\affiliation{Department of Physics, Massachusetts Institute of Technology, Cambridge, Massachusetts 02139, USA}
\date{\today}

\begin{abstract}
Weyl points and line nodes are three-dimensional linear point- and line-degeneracies between two bands. In contrast to Dirac points, which are their two-dimensional analogues, Weyl points are stable in the momentum space and the associated surface states are predicted to be topologically non-trivial. However, Weyl points are yet to be discovered in nature. Here, we report photonic crystals, based on the double-gyroid structures, exhibiting frequency-isolated Weyl points with intricate phase diagrams. The surface states associated with the non-zero Chern numbers are demonstrated.
Line nodes are also found in similar geometries; the associated surface states are shown to be flat bands. Our results are readily experimentally realizable at both microwave and optical frequencies.

\end{abstract}


\maketitle

Two-dimensional (2D) electrons and photons at the energies and frequencies of Dirac points exhibit extraordinary features~\cite{Haldane:2008-PRL,Sepkhanov2007DiracTransmission,Zhang2008Zitterbewegung,Jorge2012-DiracLaser,Khanikaev-2012-Bianisotropic-QSH}.
As the best example, almost all the remarkable properties of graphene are tied to the massless Dirac fermions at its Fermi level~\cite{Neto2009graphene,Bonaccorso2010graphene}.
Topologically~\cite{Hasan2010TopologicalReview}, Dirac cones are not only the critical points for 2D phase transitions but also the unique surface manifestation of a topologically gapped 3D bulk. 
In a similar way, it is expected that if a material could be found that exhibits a 3D linear dispersion relation, it would also display a wide range of interesting physics phenomena. The associated 3D linear point-degeneracies are called ``Weyl points".
In the past year, there have been a few studies of Weyl fermions in electronics~\cite{Wan2011Weyl,Burkov2011WeylMultilayer,Halasz2011Time,Xu2011ChernSemimetal,Fang2012-Multi-Weyl,Young20123D-Dirac}. The associated Fermi-arc surface states, quantum-Hall-effect~\cite{Yang2011QHEWeyl}, novel transport properties~\cite{Hosur2012-WeylTransport} and the realization of the Adler-Bell-Jackiw anomaly~\cite{Aji2012ABJ-Weyl} are also expected. However, no observation of Weyl points has been reported.
Here, we present a theoretical discovery and detailed numerical investigation of frequency-isolated Weyl points in perturbed double-gyroid(DG) photonic crystals(PhCs) along with their complete phase diagrams and their topologically-protected surface states. PhCs containing frequency-isolated linear line-degeneracies, known as ``line nodes", and their flat-band surface states are also presented. Unlike the proposed Weyl points in electronic system thus far, our predictions in photonics are readily realizable in experiments.

Before proceeding, we first point out one intriguing distinction between the 2D Dirac points and the 3D Weyl points.
2D Dirac cones are not robust; they are only protected by the product of time-reversal-symmetry(T) and parity(P, inversion).
In 2D, Dirac cone effective Hamiltonian takes the form of $H(\mathbf{k})=v_xk_x\sigma_x+v_zk_z\sigma_z$; this form is protected by PT (product of P and T) which requires $H(\mathbf{k})$ to be real.
Thus, one can open a gap in this dispersion relation upon introducing a perturbation proportional to $\sigma_y=\left(\begin{array}{cc}0&-i\\i&0\end{array}\right)$ that is imaginary; for example, even an infinitesimal perturbation that breaks just P or just T will open a gap.
In contrast, 3D Weyl points are topologically protected gapless dispersions robust against \emph{any} perturbation.
In 3D, Weyl point dispersions are governed by the Weyl Hamiltonian $H(\mathbf{k})=v_xk_x\sigma_x+v_yk_y\sigma_y+v_zk_z\sigma_z$; the $\sigma_y$ term can exist only when PT is broken: this is a necessary condition for the existence of Weyl points. Since all three Pauli matrices are used in the Hamiltonian, one has no possibility of constructing a term that can open a gap in this two-band-degeneracy of 3D periodic systems, thereby making a single Weyl point absolutely robust to perturbations. The only way to eliminate and create Weyl points is through pair-annihilations and pair-generations of Weyl points of opposite chiralities, which typically requires a strong perturbation.

\begin{figure}[!ht]
\includegraphics[width=0.5\textwidth]{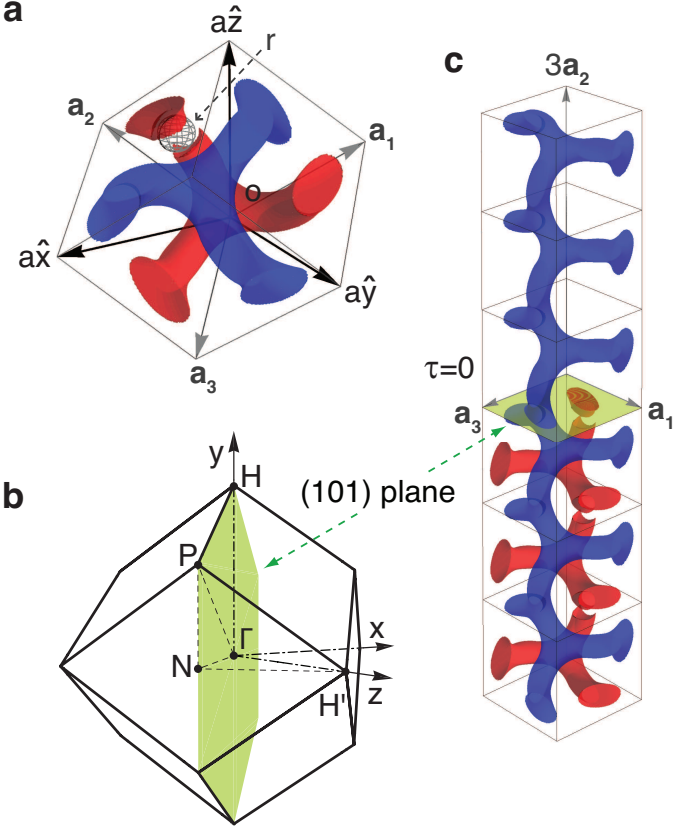}
\caption{\textbf{Real-space unit cell and reciprocal-space BZ of the 3D DG PhC.} \textbf{a}, Real space geometry in a bcc unit cell where $\mathbf{a_1}=(-1,1,1)\frac{a}{2}$, $\mathbf{a_2}=(1,-1,1)\frac{a}{2}$ and $\mathbf{a_3}=(1,1,-1)\frac{a}{2}$. The two identical gyroid structures of red and blue colors are high refractive index ($n=4$) materials; they are inversion pairs of each other with respect to the origin($\textrm{o}$). An air sphere ($r/a=0.13$) located at $(\frac{1}{4},-\frac{1}{8},\frac{1}{2})a$ breaks the inversion symmetry of the system. 
\textbf{b}, The BZ of the bcc lattice. Weyl points and linear line-degeneracies of investigation in this letter always lie in the green plane through the origin($\Gamma$). $\Gamma$-N is along $[\overline{1}01]$ and $\Gamma$-H is along $[010]$($\hat{y}$).
\textbf{c}, An air-isolated DG surface can be formed by terminating the perturbed gyroid (red) but not the other (blue). The SG PhC on the top has a large complete bandgap as shown in Fig. \ref{Fig:2_bands}\textbf{a}.}
\label{Fig:1_bccBZ}
\end{figure}

\begin{figure*}[!ht]
\includegraphics[width=1.0\textwidth]{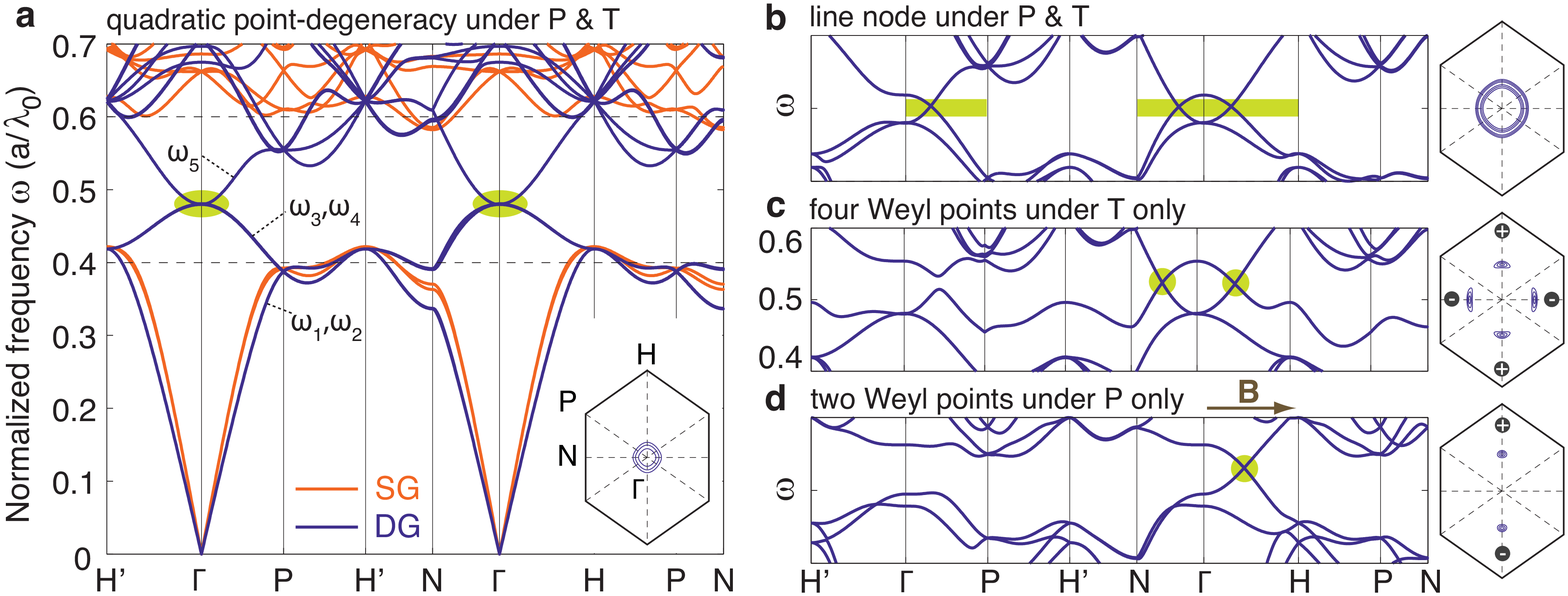}
\caption{\textbf{Gapless photonic bandstructures of the DG PhCs without (\textbf{a}) and with perturbations (\textbf{b,c,d}).} 
\textbf{a}, The original DG PhC bandstructure has a three-fold degeneracy at $\Gamma$ among the 3$^{\textrm{rd}}$, 4$^{\textrm{th}}$ and 5$^{\textrm{th}}$ bands in a pseudo-gap. The SG PhC has a huge frequency gap covering the interesting frequency region of the DG PhC dispersion. 
\textbf{b}, We place the two air-spheres on the two gyroids: one air-sphere ($r/a=0.07$) is located at $(\frac{1}{4},-\frac{1}{8},\frac{1}{2})a$; the other is its inversion symmetric pair. Under this perturbation, the 4$^{\textrm{th}}$ and 5$^{\textrm{th}}$ bands linearly touch in a closed line around $\Gamma$ in the $\Gamma$-N-P-H plane. The linear crossing line is highlighted by a green stripe; this structure does not exhibit any Weyl points yet.
\textbf{c}, We apply a P-breaking perturbation by placing only one air sphere ($r/a=0.10$) in one of the gyroids but not the other. Two pairs of Weyl points appear(highlighted by the green circles): one pair appears along $\Gamma$-H and the other pair appears along $\Gamma$-N. 
\textbf{d}, We apply a T-breaking perturbation (P conserves) by applying DC magnetic field (dimensionless $|\mathbf{B}|=0.875$) on the DG PhC without air-spheres. Only one pair of Weyl points appears (highlighed by the green circle) along the direction of the magnetic field ($\Gamma$-N). Since P and T are not broken at the same time in these PhCs, \textbf{k} and -\textbf{k} are degenerate in the bandstructures. A few lowest-value equi-frequency contours of the frequency difference between the 4th and 5th bands are shown (hexagonal insets) for each bandstructure in the $\Gamma$-N-P-H plane; the contour spacing is 0.004 in normalized frequency; ``+" and ``-" are used to label the chiralities of the Weyl points.
All the dispersion behaviors close to the degeneracy points in this figure can be well described by the low-energy-theory model in the supplementary material.
}
\label{Fig:2_bands}
\end{figure*}

Our starting point for obtaining Weyl points is a PhC whose bandstructure exhibits a three-fold degeneracy(quadratic in all three directions) at the BZ center ($\Gamma$) in an otherwise complete bandgap.
This pseudo-gap system was reported in a DG PhC~\cite{Maldovan2002DG}, whose real space structure is shown in Fig. \ref{Fig:1_bccBZ}\textbf{a}. When the air-sphere is not introduced, its gapless bandstructure is plotted in Fig. \ref{Fig:2_bands}\textbf{a}\footnote{The calculations in this paper were performed using the MIT Photonic-Bands (MPB) package.}. As shown below, applying different perturbations to this structure will lead us to PhCs with frequency-isolated linear point and line degeneracies (i.e. Weyl points and line nodes).

A DG PhC consists of two single gyroids (SG) in a body-centered-cubic(bcc) lattice. An SG surface is a bi-continuous triply-periodic minimal surface defined by $g(\mathbf{r})=\sin(2\pi x)\cos(2\pi y)+\sin(2\pi y)\cos(2\pi z)+\sin(2\pi z)\cos(2\pi x)$~\cite{Wohlgemuth2001Triply}. The red gyroid in Fig. \ref{Fig:1_bccBZ}\textbf{a} is defined by filling the inner space of the isosurface($g(\mathbf{r})>1.1$) to be high refractive-index material($\sqrt{\epsilon}=n=4$)  and air otherwise; magnetic permeability $\mu$ is unity everywhere. (This setting corresponds to Germanium and air at optical frequencies.) The blue gyroid is the inversion pair of the red gyroid with respect to the origin; the two gyroids do not overlap in space. The bandstructures of both SG PhC and DG PhC are plotted in Fig. \ref{Fig:2_bands}\textbf{a} by orange and blue colors respectively. The SG PhC has a 32\% complete bandgap between the $2^{\textrm{nd}}$ and $3^{\textrm{nd}}$ bands from 0.42 to 0.58 in normalized frequencies~\cite{Maldovan2002DG}. The space group of the SG is $I4_132$ (No. 214)~\cite{Manes2011symmetry,hahn2002international}; the DG structure belongs to $Ia\overline{3}d$ (No. 230), which is a direct-product group of $I4_132$ and inversion. Both of the space groups are non-symmorphic and have three-dimensional representations~\cite{Aroyo:xo5013}. This explains the three-fold degeneracy in the DG PhC bandstructure, among 3$^{\textrm{rd}}$, 4$^{\textrm{th}}$ and 5$^{\textrm{th}}$ bands at the center of the BZ ($\Gamma$), highlighted with green ellipses in Fig. \ref{Fig:2_bands}\textbf{a}. The 1$^{\textrm{st}}$ and 2$^{\textrm{nd}}$ bands are almost degenerate; so are the 3$^{\textrm{rd}}$ and 4$^{\textrm{th}}$ bands: they concave down and touch the 5$^{\textrm{th}}$ band which concaves up. The three-fold degenerate point is well isolated in frequency from other parts of the dispersion diagram of the DG bandstructure, making it an ideal starting point for applying symmetry-breaking perturbations.

\begin{figure*}[!ht]
\includegraphics[width=0.9\textwidth]{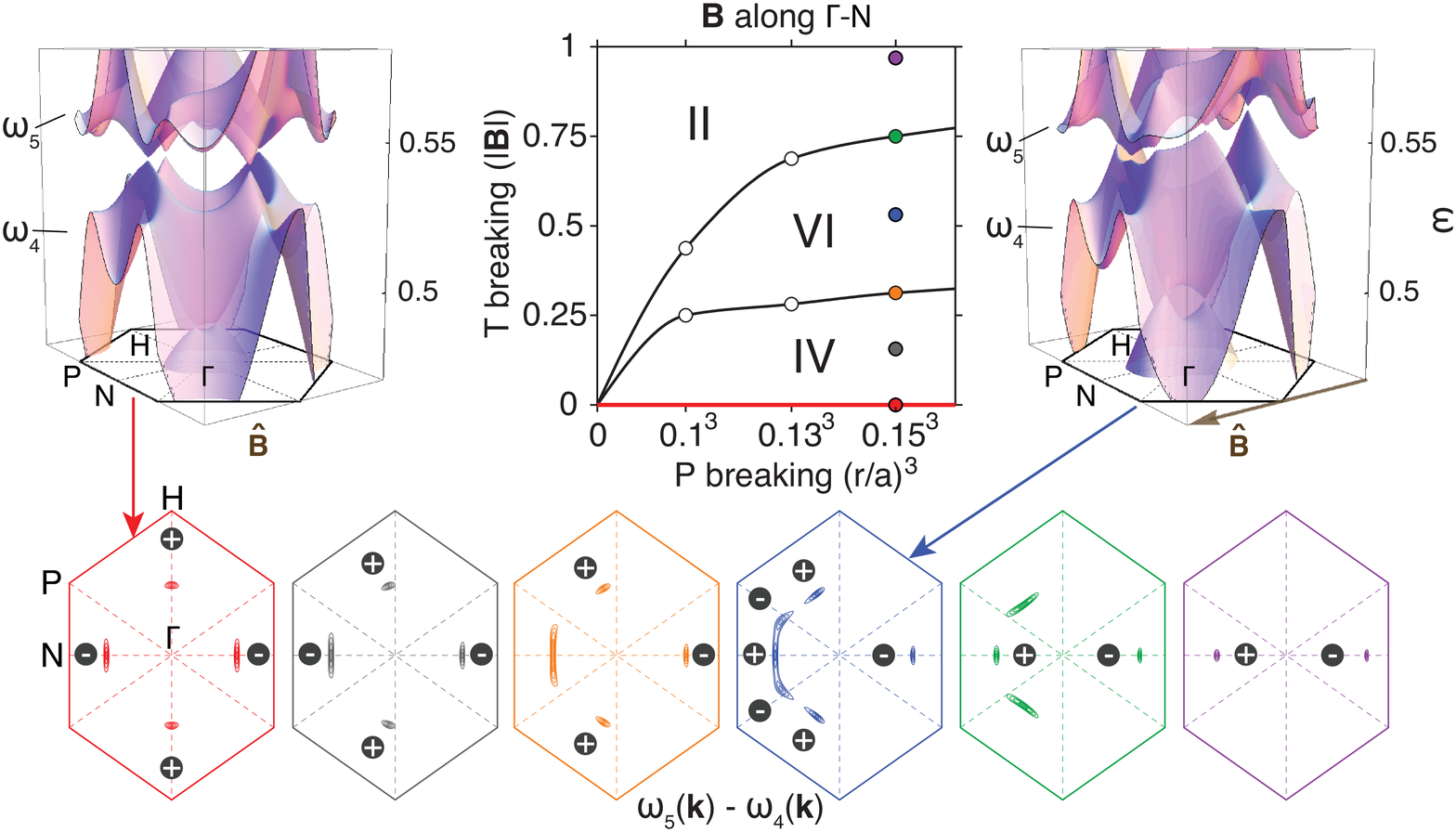}
\caption{\textbf{Phase diagram of the Weyl points when the magnetic field is applied along $\Gamma$-N.}
Each phase is labeled by the number(Roman numerals) of Weyl points. The phase transition line is found by scanning the $\mathbf{B}$ field amplitude in the calculation for three different P-breaking strengths. The pure P-breaking phase (x-axis) is highlighted in red; it is realizable at the optical frequencies with essentially lossless materials. Six 2D contour plots show the locations of the Weyl points at representative positions in the phase diagram; the contour spacing is 0.002 in normalized frequency. The contour plots share the same color as the circles of their corresponding positions in the phase diagram. 
Two surface dispersion plots show the entire band dispersions $\omega(\textbf{k})$ of the 3$^{\textrm{rd}}$, 4$^{\textrm{th}}$ and 5$^{\textrm{th}}$ bands for two particular cases, where $\textbf{k}$ lies in the green plane of Fig. \ref{Fig:1_bccBZ}\textbf{b}.}
\label{Fig:4_B-101}
\end{figure*}

The three-fold degeneracy of quadratic dispersions at $\Gamma$ can be lifted by breaking the $I4_132$ space group without breaking P or T symmetries. This is done by replacing a part of the gyroid material with two air-spheres (one on each gyroid). The first air-sphere is placed in the red gyroid at $(\frac{1}{4},-\frac{1}{8},\frac{1}{2})a$, as illustrated in Fig. \ref{Fig:1_bccBZ}\textbf{a}; and the other is its inversion pair in the blue gyroid (not illustrated in Fig. \ref{Fig:1_bccBZ}\textbf{a}). This perturbation lifts the $5^{\textrm{th}}$ band out of the three-fold degeneracy with the $3^{\textrm{rd}}$ and $4^{\textrm{th}}$ bands at $\Gamma$, as shown in Fig. \ref{Fig:2_bands}\textbf{b}. The $4^{\textrm{th}}$ and $5^{\textrm{th}}$ bands linearly cross each other, forming a closed line-degeneracy around the $\Gamma$ point in the $\Gamma$-N-P-H plane, inside an otherwise complete frequency gap. It is worth pointing out that this bandstructure, although not exhibiting Weyl points, is interesting in itself in analogy to the line-node semimetals~\cite{Burkov2011nodal}. We show its flat surface dispersions towards the end of this paper.

In what follows, we break the PT symmetry to obtain Weyl points of photons for the first time.
We start with the structure from Fig. \ref{Fig:2_bands}\textbf{a}. First, we break P while preserving T. Since T maps a Weyl point at $\mathbf{k}$ to $-\mathbf{k}$ with the same chirality, there must exist at least two other Weyl points, both of opposite chirality, to neutralize the whole system. So the minimal number of Weyl points in this case has to be four. We break P by placing only one air sphere on one of the gyroids (but not the other) at $(\frac{1}{4},-\frac{1}{8},\frac{1}{2})a$\footnote{The SG and DG structures are invariant under $C_2$ rotation along $[\overline{1}01]$($\Gamma$-N) at $(\frac{1}{4},-\frac{1}{8},\frac{1}{2})a$.}, as illustrated in Fig. \ref{Fig:1_bccBZ}(\textbf{a}). Under this pure P-breaking perturbation\footnote{The continuous structural changes for breaking inversion is an advantage of PhCs over electronic crystals in studying Weyl points.}, two pairs of Weyl points,  shown in Fig. \ref{Fig:2_bands}\textbf{c}, emerge along $\Gamma$-N and $\Gamma$-H directions. The fact that all the Weyl points appear along high-symmetry lines significantly simplifies the analysis. There are no other states in the vicinity of the Weyl points' frequencies.

Second, DC magnetic fields (\textbf{B}), along different directions, are applied to the original DG PhC structure in Fig. \ref{Fig:2_bands}\textbf{a} to break the T while preserving P. We assume the high-index gyroid material is gyroelectric and use a generic model~\cite{Bouchaud1989ferrimagnets} to describe its magnetic response. When \textbf{B} is along $\hat{z}$, we assume the permittivity tensor takes the form of
\begin{eqnarray}
\epsilon(|\mathbf{B}|)=\left(\begin{array}{ccc}\epsilon_{11}(|\mathbf{B}|)&i\epsilon_{12}(|\mathbf{B}|)&0\\-i\epsilon_{12}(|\mathbf{B}|)&\epsilon_{11}(|\mathbf{B}|)&0\\0&0&\epsilon\end{array}\right)
\end{eqnarray}
where $\textrm{det}(\epsilon(|\mathbf{B}|))=(\epsilon_{11}^2(|\mathbf{B}|)-\epsilon_{12}^2(|\mathbf{B}|))\epsilon=\epsilon^3$; this constant determinant condition ensures the dispersions as a whole do not move much in frequency with the external DC \textbf{B} fields. The dimensionless effective magnetic field intensity is defined as $|\mathbf{B}|\equiv\epsilon_{12}/\epsilon$ in this paper. When \textbf{B} field is along other directions, the corresponding $\epsilon$ tensor can be obtained through coordinate transformations. (Note the T-breaking can be equally well implemented via $\mu$ for gyromagnetic materials~\cite{Wang2009}.) Under this pure T-breaking perturbation, only a single pair of Weyl points emerges along the direction of the magnetic field. This is the minimum number of Weyl points that can exist under the inversion symmetry. These two Weyl points are still frequency-degenerate: P maps a Weyl point at $\mathbf{k}$ to $-\mathbf{k}$ with the opposite chirality. An example of this is shown in Fig. \ref{Fig:2_bands}\textbf{d}.

Third, we apply both P and T breaking perturbations at the same time to observe the phase transitions between the two(II) Weyl points in the pure T-breaking phase and the four(IV) Weyl points in the pure P-breaking phase. Interestingly enough, different magnetic field directions produce strikingly different phase diagrams. 
When \textbf{B} is applied along $\Gamma$-H, only two phases exist: the T-breaking dominated phase (II) and the P-breaking dominated phase (IV). 
The pure P-breaking phase, shown in the contour plot Fig. \ref{Fig:2_bands}\textbf{c}, has four Weyl points: two with positive chiralities along $\Gamma$-H and two with negative chiralities along $\Gamma$-N. Applying magnetic field along the $\Gamma$-H direction drives the two negative-chirality Weyl points towards the positive one along $\Gamma$-$\overline{\textrm{H}}$. 
Increasing the magnetic field further annihilates two of the Weyl points of opposite chiralities and the system enters the T-breaking dominated phase which has only two Weyl points along $\Gamma$-H. An example where only T is broken is shown in Fig. \ref{Fig:2_bands}\textbf{d}.
Detailed description of this phase diagram is shown in the supplementary Fig. A.

An even richer phase diagram, shown in Fig. \ref{Fig:4_B-101}, appears when we switch the magnetic field to the $\Gamma$-N direction. The system undergoes two phase transitions from the P-breaking dominated phase(IV) to the T-breaking dominated phase(II). The extra intermediate phase(VI) of six Weyl points is generated when one of the negative-chirality Weyl points in the pure P-breaking phase (red contour plot) splits (orange contour plot) into three Weyl points (blue contour plot) under the increasing $\textbf{B}$ field. Among these three Weyl points, the original Weyl point flips its chirality from negative to positive; the other two negative Weyl points move away from their creation position towards the neighboring two positive ones. The two pairs of Weyl points eventually meet (green contour plot) and annihilate simultaneously, leaving the remaining two Weyl points (purple contour plot) along the direction of the magnetic field($\Gamma$-N). When two Weyl points annihilate, the band dispersions are quadratic along the directions in which they meet.

In general, P-T phase diagrams depend on the form of the perturbations: the DC magnetic field can be oriented along an arbitrary direction; in addition, there are many different ways to break the inversion symmetry.

\begin{figure}[!ht]
\includegraphics[width=0.45\textwidth]{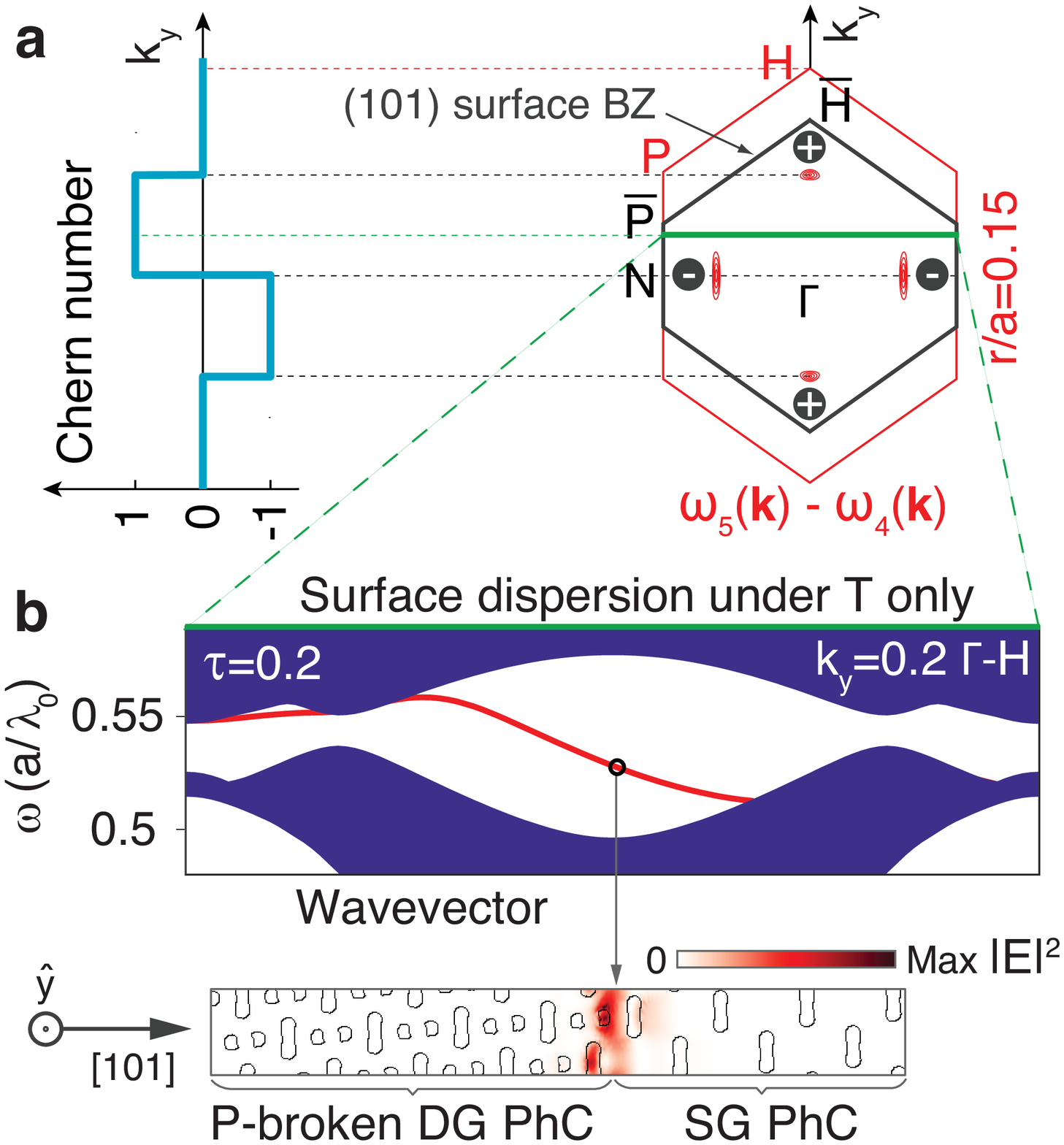}
\caption{\textbf{Topological surface states of a Weyl-point PhC.} \textbf{a}, Illustration of the Chern number of the lower bulk bands on the plane moving along $\hat{k_y}$ in the 3D bulk BZ. \textbf{b}, An example of a topologically protected surface state of a Weyl point PhC. The projected bulk states are plotted in blue in the surface bandstructure; the surface state dispersion relation is plotted in red. The spatial field distribution ($|\mathbf{E}|^2$) of one surface mode, at the center of the $\mathbf{k}$ axis, is plotted with the contour outlines of the refractive index profile. The surface state is localized at the interface between the DG PhC and the SG PhC.}
\label{Fig:surface}
\end{figure}

Weyl points are topologically stable objects in the 3D BZ: they act as monopoles of Berry flux in momentum space, and hence are intimately related to the topological invariant known as the Chern number~\cite{Raghu2008QHE}.
The Chern number can be defined for a single bulk band or a set of bands, where the Chern numbers of the individual bands are summed, on any closed 2D surface in the 3D BZ.
The difference of the Chern numbers defined on two surfaces, of all bands below the Weyl point frequencies, equals the sum of the chiralities of the Weyl points enclosed in between the two surfaces~\cite{Yang2011QHEWeyl}.
In particular, we illustrate in Fig. \ref{Fig:surface}\textbf{a}, the Chern numbers of the 2D planes perpendicular to $\Gamma$-H($\hat{y})$; they are closed surfaces in the 3D periodic BZ.
The first contour plot (red), of a pure P-breaking phase in Fig. \ref{Fig:4_B-101}, is used as an example.
The Chern number of all lower bands on the plane is plotted in light blue on the left side. The Chern number vanishes when the plane is outside of the Weyl points.
It changes by one when the plane moves across one Weyl point and changes by two when the plane crosses two Weyl points of the same chirality.

As a result of the non-zero Chern number of all the lower bulk bands (1$^\textrm{st}$, 2$^\textrm{nd}$, 3$^\textrm{rd}$ and 4$^\textrm{th}$), there exists topologically-protected gapless chiral surface states inside the bandgap (between 4$^\textrm{th}$ and 5$^\textrm{th}$ bands) away from the Weyl points~\cite{Wang2008PRL}. In order to prevent the surface states from radiating into air, we interface the DG PhC with an SG PhC by removing only the perturbed gyroid above the (101) surface: this is illustrated in Fig. \ref{Fig:1_bccBZ}\textbf{c}. The surface states of finite $k_y$, are trapped by the partial gap of the DG PhC and the full 3D gap of the SG PhC. We define a termination parameter $\tau$($0\le\tau<1$) to indicate the periodically equivalent termination positions along the [101] direction. $\tau=0$ is set at the origin of the unit cell as shown in Fig. \ref{Fig:1_bccBZ}\textbf{c}. An example of the non-trivial gapless surface dispersion, along a line cut in the 2D surface BZ in Fig. \ref{Fig:surface}\textbf{a}, is plotted in Fig. \ref{Fig:surface}\textbf{b}. A surface mode profile is also shown. Since T is not broken in this example, the surface dispersions are degenerate between $\mathbf{k}$ and $-\mathbf{k}$, while Chern number flips sign under T symmetry. This means the surface modes have opposite chiralities when travelling along $+\hat{y}$ and $-\hat{y}$ directions at the same frequency. The above analysis on Chern numbers and surface modes applies for states along other directions and for other Weyl phases as well.
We stress that the non-zero Chern numbers and the associated chiral surface states in this example are obtained without breaking T~\cite{Haldane:2008-PRL,Wang2009}; thus, this phenomenon is readily realizable at optical frequencies.

\begin{figure}[!ht]
\includegraphics[width=0.5\textwidth]{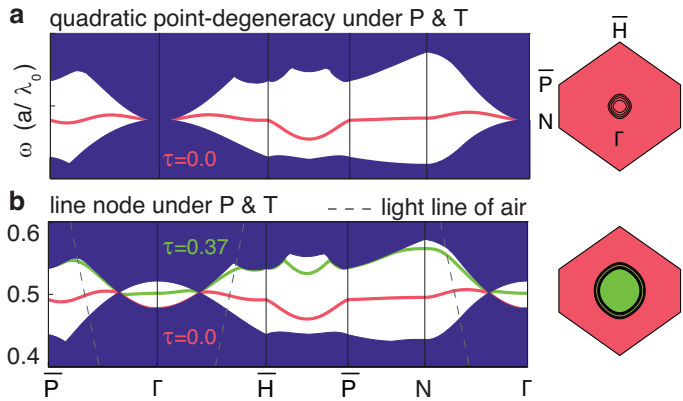}
\caption{\textbf{Flat surface dispersions of two pseudo-gapped PhCs without Weyl points.} \textbf{a}, Surface dispersion at the interface between the SG and the original DG of a bulk quadratic point-degeneracy as in Fig. \ref{Fig:2_bands}\textbf{a}. \textbf{b}, Two surface dispersions at the interface between the SG and the DG modified by two air spheres as in Fig. 
\ref{Fig:2_bands}\textbf{b}. They are flat in two different regions of the surface BZ separated by the linear line-degenerate bulk states. The surface BZ is colored in the region where the surface states are flat.}
\label{Fig:flatsurf}
\end{figure}

The surface states of the other two gapless PhCs in Fig. \ref{Fig:2_bands}\textbf{a} and \textbf{b} do not involve non-zero Chern numbers, but they can exhibit very flat bands which is also an interesting feature. Fig. \ref{Fig:flatsurf}\textbf{a} shows one surface state of the DG PhC of quadratic point-degeneracy in the bulk pseudo-gap. When $\tau$ varies periodically, the surface dispersion, at every surface \textbf{k} point, moves from the air-band(conduction band) through the pseudo-gap to the dielectric-band(valence band)~\cite{Meade1991}. At the center of the BZ, the surface dispersion is pinned into the bulk states at the degeneracy point. The surface dispersion of $\tau=0.0$, apart from the $\overline{\textrm{H}}$-$\overline{\textrm{P}}$ region, is very flat. The high density of states associated with the flat surface dispersion can be potentially useful for enhancing the efficiency of light-emitters at the surface.
Even more interesting surface states are shown in Fig. \ref{Fig:flatsurf}\textbf{b} for the line-node PhC. The nodal line bulk states project onto the surface BZ as a closed line which separates the surface BZ into two disconnected areas. So the surface dispersions can be flat bands in these two separate regions of the BZ. The green dispersion in Fig. \ref{Fig:flatsurf}\textbf{b} has all its frequencies nearly degenerate inside the line-node area; while the red dispersion is relatively flat in the rest of the BZ.
The general features of the flat surface dispersions do not change when the line-node PhC is terminated by other means. For example, when the SG PhC is replaced by air, one could selectively enhance, by changing the surface termination, the light emission of surface sources into either radiative or non-radiative modes.

In conclusion, this letter provides a detailed design of the minimum numbers of frequency-isolated Weyl points and line nodes along the high-symmetry lines and planes in the BZ of the DG PhCs. Complete P-T phase diagrams are investigated through annihilations and creations of the Weyl pairs. The topologically non-trivial surface states of Weyl-point PhCs and flat surface dispersions of line-node PhCs are presented. This work potentially opens doors to new paradigms in photonics: topologically-protected chiral surface states realizable at optical frequencies for the first time, radiation-controllable photonic density of states enhanced by flat surface dispersions, novel transmission properties of light in the vicinity of the degeneracy frequencies, and other unconventional phenomena associated with the density-of-states and dispersions of the close to the degeneracy points that were not available before. 
Experimentally, our design can be realized with many available techniques. Germanium can either be directly synthesized~\cite{Armatas2006Ge} into DG strucutures or backfilled~\cite{Garcia2007GeWoodpile} into DG polymer templates made by block-copolymers~\cite{Saranathan2010PNASdg} or interference lithography~\cite{Ullal:03}. The templates for geometrically-perturbed DGs can be made by 3D lithography or 3D printing. All presented results without magnetic fields are readily realizable with Germanium at optical frequencies; Weyl phases involving T-breaking can be readily demonstrated with gyromagnetic materials at microwave frequencies~\cite{Wang2009}.

We would like to thank Fa Wang, Martin Maldovan, Zheng Wang, Steven G. Johnson, Ashvin Vishwanath and Dung-Hai Lee for helpful discussions.
This work was supported in part by the U.S.A.R.O. through the ISN, under Contract No.W911NF-07-D-0004.
L.L. was supported in part by the MRSEC Program of the NSF under Award No. DMR-0819762. 
M.S. and L.L. were supported in part by the MIT S3TEC Energy Research Frontier Center of the Department of Energy under Grant No. DE-SC0001299.
L.F. was supported by the start-up funds from MIT.


\end{document}